\documentclass{aa} 

\usepackage{graphicx}
\usepackage{ulem}
\usepackage[T1]{fontenc}
\usepackage{tabularx}
\usepackage{ulem}

\usepackage[colorlinks=true, allcolors=blue]{hyperref}
\hypersetup{
     colorlinks   = true,
     citecolor    = blue,
     urlcolor = blue
}
\usepackage{txfonts}
\usepackage{caption}

\DeclareCaptionFormat{cont}{#1 (cont.)#2#3\par}
\begin{document}

   \title{The Northern Cross Fast Radio Burst project}

   \subtitle{IV. Multi-wavelength study of the actively repeating FRB 20220912A}

   \author{D. Pelliciari
          \inst{1,2}
          \and
          G. Bernardi\inst{1,3,4}
          \and
          M. Pilia\inst{5}
          \and
          G. Naldi\inst{1}
          \and
          G. Maccaferri\inst{1}
          \and
          F. Verrecchia\inst{14,15}
          \and
          C. Casentini\inst{11}
          \and
          M. Perri\inst{14,15}
          \and
          F. Kirsten\inst{6,7}
          \and
          G. Bianchi\inst{1}
          \and
          C.~Bortolotti\inst{1} 
          \and
          L. Bruno\inst{1,2}
          \and
          D.~Dallacasa\inst{1,2} 
          \and
          P.~Esposito\inst{9}
          \and
          A.~Geminardi\inst{5,9,10}
          \and
          S.~Giarratana\inst{1,2}
          \and
          M.~Giroletti\inst{1}
          \and
            R.~Lulli\inst{1}
          \and
            A.~Maccaferri\inst{1}
          \and
            A.~Magro\inst{11}
          \and
            A.~Mattana\inst{1}
          \and
            F.~Perini\inst{1}
          \and
          G.~Pupillo\inst{1}
          \and
            M.~Roma\inst{1}
          \and
            M.~Schiaffino\inst{1}
          \and
          G.~Setti\inst{1,2}
          \and
            M.~Tavani\inst{12,13} 
          \and
          M.~Trudu\inst{5}
          \and A.~Zanichelli\inst{1}
          }

   \institute{INAF-Istituto di Radio Astronomia (IRA), via Piero Gobetti 101, Bologna, Italy\\
              \email{davide.pelliciari@inaf.it}
         \and
             Dipartimento di Fisica e Astronomia, Universit\'{a} di Bologna, via Gobetti 93/2, 40129 Bologna, Italy
         \and
             South African Radio Astronomy Observatory, Black River Park, 2 Fir Street, Observatory, Cape Town, 7925, South Africa
         \and
             Department of Physics and Electronics, Rhodes University, PO Box 94, Makhanda, 6140, South Africa
         \and
             INAF-Osservatorio Astronomico di Cagliari, via della Scienza 5, I-09047, Selargius (CA), Italy
        \and
            Department of Space, Earth and Environment, Chalmers University of Technology, Onsala Space Observatory, 439 92, Onsala, Sweden
        \and ASTRON, Netherlands Institute for Radio Astronomy, Oude Hoogeveensedijk 4, 7991 PD Dwingeloo, The Netherlands
        \and
            INAF-Osservatorio di Astrofisica e Scienza dello Spazio di Bologna, via Piero Gobetti 93/3, 40129, Bologna, Italy
        \and
            Scuola Universitaria Superiore IUSS Pavia, Piazza della Vittoria 15, 27100 Pavia, Italy
        \and
            Department of Physics, University of Trento, via Sommarive 14, 38123 Povo (TN), Italy
        \and
            Institute of Space Sciences and Astronomy (ISSA), University of Malta, Msida, MSD 2080, Malta
        \and
            INAF/IAPS, via del Fosso del Cavaliere 100, I-00133 Roma (RM), Italy
        \and
            Università degli Studi di Roma "Tor Vergata", via della Ricerca Scientifica 1, I-00133 Roma (RM), Italy
        \and
            SSDC/ASI, via del Politecnico snc, I-00133 Roma (RM), Italy
        \and
            INAF-Osservatorio Astronomico di Roma, via Frascati 33, 00078 Monte Porzio Catone (RM), Italy}

   \date{XXX-XXX-XXX}

 
  \abstract
   {Fast radio bursts (FRBs) are energetic, millisecond-duration radio pulses observed at extragalactic distances and whose origins are still a subject of heated debate. A fraction of the FRB population have shown repeating bursts, however it's still unclear whether these represent a distinct class of sources.}
   {We investigated the bursting behaviour of FRB 20220912A, one of the most active repeating FRBs known thus far. In particular, we focused on its burst energy distribution, linked to the source energetics, and its emission spectrum, with the latter directly related to the underlying emission mechanism.
   }
   {We monitored FRB 20220912A at $408$ MHz with the Northern Cross radio telescope and at $1.4$ GHz using the $32$-m Medicina Grueff radio telescope. Additionally, we conducted $1.2$ GHz observations taken with the upgraded Giant Meter Wave Radio Telescope (uGMRT) searching for a persistent radio source coincident with FRB 20220912A, and we included high energy observations in the $0.3$--$10$ keV, $0.4$--$100$ MeV and $0.03$--$30$ GeV energy range.
   } 
   {We report 16 new bursts from FRB 20220912A at $408$ MHz during the period between October 16$^{\rm th}$ 2022 and December 31$^{\rm st}$ 2023. Their cumulative spectral energy distribution follows a power law with slope $\alpha_E = -1.3 \pm 0.2$ and we measured a repetition rate of $0.19 \pm 0.03$ hr$^{-1}$ for bursts having a fluence of $\mathcal{F} \geq 17$ Jy ms. Furthermore, we report no detections at $1.4$ GHz for $\mathcal{F} \geq 20$ Jy ms. These non-detections imply an upper limit of $\beta < -2.3$, with $\beta$ being the 408 MHz -- 1.4 GHz spectral index of FRB 20220912A. This is inconsistent with positive $\beta$ values found for the only two known cases in which an FRB has been detected in separate spectral bands. We find that FRB 20220912A shows a decline of four orders of magnitude in its bursting activity at $1.4$ GHz over a timescale of one year, while remaining active at $408$ MHz. The cumulative spectral energy distribution (SED) shows a flattening for spectral energy $E_\nu \geq 10^{31}$ erg Hz$^{-1}$, a feature seen thus far in only two hyperactive repeaters. In particular, we highlight a strong similarity between FRB 20220912A and FRB 20201124A, with respect to both the energy and repetition rate ranges. We also find a radio continuum source with $240 \pm 36$~$\mu$Jy flux density at 1.2~GHz, centered on the FRB 20220912A coordinates. Finally, we place an upper limit on the $\gamma$ to radio burst efficiency $\eta$ to be $\eta < 1.5 \times 10^9$ at 99.7\% confidence level, in the $0.4$ -- $30$ MeV energy range.
   }
   {The strong similarity between the cumulative energy distributions of FRB 20220912A and FRB 20201124A indicate that bursts from these sources are generated via similar emission mechanisms. Our upper limit on $\beta$ suggests that the spectrum of FRB 20220912A is intrinsically narrow-band. The radio continuum source detected at 1.2~GHz is likely due to a star formation environment surrounding the FRB, given the absence of a source compact on millisecond scales brighter than 48~$\mu$Jy beam$^{-1}$ (cit). Finally, the upper limit on the ratio between the $\gamma$ and radio burst fluence disfavours a giant flare origin for the radio bursts unlike observed for the Galactic magnetar SGR 1806-20.
   }

   
   \titlerunning{Multi-wavelength study of FRB 20220912A}

   \maketitle

\section{Introduction}\label{sec:Intro}
Significant observational and theoretical efforts have been undertaken to understand the origin of millisecond-long radio flashes of extragalactic nature, known as fast radio bursts \citep[FRBs; see][for recent reviews]{Bailes22, Petroff22, Zhang22_rev}. A number of models invoke magnetars \citep[e.g.][]{DuncanThompson, ThompsonDuncan}, namely neutron stars (NSs) powered by the decay of strong ($10^{14} - 10^{16}$ G) magnetic fields, as FRB progenitors \citep{Popov13, Beloborodov19, Liubarsky20, Lu20, Bochenek21, Sobacchi22}. This hypothesis is supported by the simultaneous detection of an FRB-like burst, FRB 20200428 \citep{CHIME20b, Bochenek20a}, with an X-ray outburst from the Galactic magnetar SGR J1935+2154 \citep{Mereghetti20, Ridnaia20, Li21, Tavani21}. 

Nowadays there are about $\sim 800$ distinct known FRB sources \citep{chimecat, Xu23} and most of them classified as one-off events. However, about $50$ sources (so-called repeaters) have shown repeated emission \citep{CHIME23}, ruling out catastrophic events as their origin. It is unclear whether all FRB sources are repeating in nature, although bursts from repeaters are statistically wider in temporal width and narrower in bandwidth compared to one-off FRBs \citep{Pleunis21}. An interesting feature that has emerged from very long monitoring of the two hyperactive repeaters FRB 20121102A (R1) and FRB 20201124A is the flattening of their burst energy distributions at the highest burst energies \citep{Hewitt22, Jahns23, Kirsten23}. This suggests a possible link between repeating and non repeating FRB sources \citep{James22b, Kirsten23}, the latter presenting a flat luminosity distribution \citep{James22, James22b}, which could potentially imply that the most energetic bursts are produced by a different emission mechanism compared to the less energetic ones. 

Among repeaters, various differences are found, especially in their observed level of activity. Indeed, the burst rate of repeaters spans a wide range of values, ranging from less active sources, which can exhibit a burst rate as low as $\sim 10^{-3}\ \rm hr^{-1}$ \citep{CHIME23} to the most active ones showing sporadic burst storms in which the repetition rate rises up to several hundreds of bursts per hour \citep{Li21, Nimmo22, Xu2022, Zhang22c, Zhang23, Feng23}. On the other hand FRB 20180916B, which shows a $16.3 \pm 2.6$ days periodic window of activity \citep{Pleunis21, PastorMarazuela21}, has not revealed any burst storms, given that its repetition rate is consistent with an origin coming from a Poissonian process \citep{Ketan23}. 

Repeating FRBs have been observed with extremely narrow spectra \citep{Kumar21, PastorMarazuela21, Pleunis21, Zhou22, Zhang23, Sheikh23}, thus hindering a multi-band detection, as well as a broad-band, simultaneous spectral index measurement, that is crucial for investigating the underlying FRB emission mechanism and to exclude progenitor models \citep[e.g.][]{BurkeSpolaor16}. However, some exceptions have been reported. Remarkably, a single burst from R1 has been simultaneously detected at $1.4$ GHz and $3$ GHz using the Arecibo radiotelescope and the Karl G. Jansky Very Large Array (VLA), respectively \citep{Law17}. Assuming a power law spectrum $F(\nu) \propto \nu^{\ \beta}$, the authors obtained a spectral index of $\beta = 2.1$. However, the latter result is inconsistent with the non-detection at $4.8$ GHz conducted simultaneously with the Effelsberg radio telescope \citep{Law17}. Therefore, the authors concluded that a single power-law function was not a good description for the broad-band spectrum of the source. Furthermore, \citet{Chawla20} reported a coincident detection of FRB 20180916B in adjacent frequency bands of the Robert C. Byrd Green Bank Telescope (GBT) ($300-400$ MHz) and the Canadian Hydrogen Intensity Mapping Experiment \citep[CHIME; $400-800$ MHz,][]{CHIMEoverview}. In this case, the burst in the CHIME band is downwardly drifting into the GBT band. This effect, known as `sad trombone', is commonly observed in the morphology of repeater bursts \citep[e.g.][]{Hessels19}. In the same work, no bursts have been detected in the Low Frequency Array (LOFAR) $110-190$ MHz band, implying a lower limit on the broad-band spectral index of $\beta > -1.0$. Finally, the simultaneous detection of FRB 20200428 at $600$ MHz \citep{CHIME20b} and $1.4$ GHz \citep{Bochenek20a} gives a rough power law broad-band spectrum of $\beta \sim 1$. However, in this case the flux density measured by CHIME is poorly constrained, given that this has been a sidelobe detection. 

\begin{figure*}
    \centering
	\includegraphics[width=1.9\columnwidth]{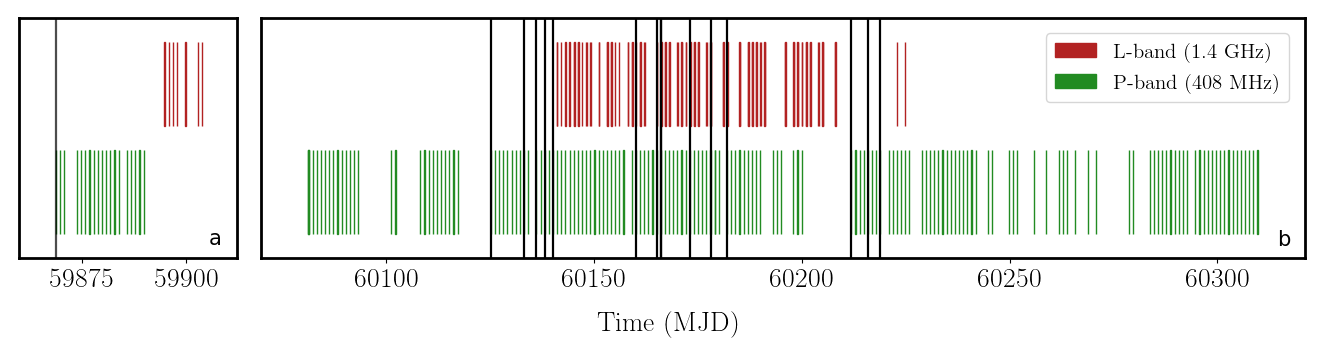}
    \caption{Overview of the FRB 20220912A monitoring campaign. Panel \textit{a} displays the MJD range $59868$ -- $59903$ ($35$ days), while panel \textit{b} shows the MJD range $60081$ -- $60309$ ($228$ days). Green (red) vertical blocks indicate observations at 408 MHz (1.4 GHz). Vertical black lines represent burst detections at 408 MHz. Each observing session at 408 MHz last $\sim 35$ minutes, while 1.4 GHz observations started $15$ minutes before the P band run and last $\sim 60$ minutes.}
    \label{obscamp}
\end{figure*}

The chance to observe FRBs more than once allowed in-depth studies of these elusive sources, and helped in their accurate association with host galaxies \citep[e.g.][]{Gordon23}. The milliarcsecond localisation precision achieved for a number of actively repeating FRBs \citep[e.g.][]{Marcote22_PRECISE} enabled the intriguing discovery of persistent radio sources (PRSs). A PRS is spatially coincident with the FRB site and it is characterised by high luminosity \citep[$L_\nu > 10^{27}$ erg s$^{-1}$ Hz$^{-1}$,][]{Law17} and compactness \citep[$< 10$ pc, ][]{Marcote17}; however, the latter is inconsistent with typical values inferred from star-formation processes. To date, only two confirmed PRSs are known, both being associated with two actively repeating FRBs: R1 \citep{Chatterjee17} and FRB 20190520B \citep[R1-twin;][]{Niu21}. These FRB sources are very similar in terms of burst activity, host galaxy properties \citep{Niu21} and very high rotation measures (RMs) \citep{Michilli18, AnnaThomas23}. In terms of spectral energy distribution (SED), the known PRSs exhibit flat radio spectra, with a spectral index $\beta \sim -0.27$ for R1 \citep{Marcote17} and $\beta \sim -0.4$ for R1-twin \citep{Niu21, Bhandari23}. In particular, the SED of the PRS associated with R1 resembles the Crab (PSR B0531+21) pulsar wind nebula, but with a magnetic field three orders of magnitude stronger to match the implied energetics of the PRS \citep{Resmi21}. For these reasons the concordance picture for the radio emission of PRSs is a strongly ionized wind nebula powered by a young actively flaring magnetar \citep{margalitmetzger18}. Interestingly, results from very long baseline interferometry (VLBI) observations of R1-twin are also consistent with a hypernebula powered by the accretion of a central compact binary system \citep{SridharMetzger22, Bhandari23}. A third putative PRS is the one associated with FRB 20201124A, another very active FRB source \citep[e.g.][]{Zhou22}. Observations conducted with the VLA revealed the presence of a persistent, extended radio source coincident with the FRB position  \citep{Piro21}. However, at the milliarcsecond scale the same radiation is completely resolved out \citep{Nimmo22}, indicating that the continuum radiation is related to extended star formation occurring in the near environment of the FRB \citep{Piro21}. 


In September 2022, CHIME discovered FRB 20220912A, a repeating FRB source having dispersion measure (DM) of $219.46$ pc cm$^{-3}$ \citep{McKinven22}, subsequently localized with arcsecond precision in the outskirts of a moderately star forming, massive galaxy at redshift $z = 0.0771$ \citep{ravi23}. Bursts were detected at different frequencies, between $408$~MHz and $2.3$~GHz \citep[see][and references therein]{Zhang23}, with a period of particularly high activity. A burst rate of $\sim 400$~hr$^{-1}$ was observed at 1.4~GHz, for a $90\%$ fluence threshold of $4$ mJy ms \citep{Zhang23}. A fraction of the observed bursts show very narrow-band spectra \citep{Zhang23} and short durations ($\sim 16\ \mu $s), the latter usually clustered in dense burst forests \citep{Hewitt23a}. The source was recently localised at RA (J2000) = $23^{\rm h}09^{\rm m}04.8988^{\rm s} \pm 0.0003^{\rm s}$, Dec (J2000) = $48^\circ42'23.908'' \pm 0.005''$\citep{Hewitt23b}, placing it closer to the centre of the host galaxy than previously suggested. Their observations also rule out the presence of a PRS down to a $\sim 20\ \mu$Jy beam$^{-1}$ level. It was argued by \citet{ravi23} that the DM contribution by the host is low ($\leq 53$ pc cm$^{-3}$). A low DM host contribution, along with an approximately zero rotation measure \citep[RM;][]{McKinven22, Zhang23, Feng23, Hewitt23a}, corroborates the hypothesis of a clean local environment \citep{Hewitt23a}. 

In this work, we report the first multi-wavelength monitoring campaign of FRB 20220912A, carried out at $408$ MHz with the Northern Cross (NC) radio telescope, at 1.4~GHz with The Medicina Grueff $32$-m single dish and at X and $\gamma$ rays with the The Neil Gehrels Swift Observatory (Swift) \citep{Gehrels04} and AGILE \citep{Tavani09} satellites. 
Furthermore, we use new deep continuum radio observations taken with the upgraded Giant Meter Wave Radio Telescope (uGMRT) at band 5 ($1.0$ -- $1.4$ GHz) to investigate the presence of a PRS in the direction of FRB 20220912A.

The paper is structured as follows. In Section \ref{sec:obs} we describe the multi-wavelength campaign conducted on FRB 20220912A. In Section~\ref{sec:results}, we describe and discuss the results of the observations. Finally, we present our conclusions in Section \ref{sec:conclusions}. 

\section{Observations}\label{sec:obs}
\subsection{Northern Cross radio telescope}
%
%

The NC radio telescope is a T-shaped transit radio telescope deployed near Medicina (Italy). The telescope operates at $408$ MHz (P band) with a $16$~MHz bandwidth, and it is undergoing an upgrade of the receiving system \citep{Locatelli20} to enable various studies, including FRB observations. The current telescope configuration has two differences with respect to observations presented in \cite{Trudu22} and \cite{Pelliciari23}. First, it doubles the collecting area, combining sixteen cylinders of the North-South arm into a single beam, whose half-power beam width is now $1.6^\circ \times 0.25^\circ$. Second, the delay correction needed to form the beam is performed at higher cadence, namely every 5~s, effectively tracking the source across the field of view. Observations are stored to disk as 16-bit SIGPROC \citep{Lorimer11} filterbank files, with a time resolution of $138.24\ \mu$s and a $14.468$ kHz frequency channel width \citep[see][for a detailed description of the system]{Locatelli20}.

We started monitoring FRB 20220912A with eight cylinders on 16 October 16 2022, forming a single beam at the source coordinates, R.A. (J2000) = $23^{\rm h} \, 09^{\rm m} \, 04.9^{\rm s}$, Dec (J2000) $=+48^\circ \, 42' \, 25.4''$ \citep{ravi23}. After the first 14 hours on-source, the observations were interrupted and resumed on 17 May 2023, when 16 cylinders were employed. Observations ended on 31 December 2023, for a total of $122$ hr on-source. Each session of observation lasted for $\sim 35$ min.  As in \cite{Trudu22} and \cite{Pelliciari23}, we performed a weekly calibration through interferometric observations of Cas~A. A summary of the conducted observations is shown in Fig. \ref{obscamp}.


\subsection{Medicina Grueff radio telescope}

Simultaneous observations were carried out at 1.4~GHz (L band) with the Medicina Grueff 32~m dish (hereafter Medicina). The total duration of the campaign was 177 hours, made of $\sim 1$ hr daily runs. We recorded the 2-bit baseband data in both circular polarisations written to disk in VDIF format \citep{Whitney10}, using the local digital baseband converter (DBBC) system \citep{Tuccari03}. Observations are centred at 1414~MHz, sampling a 128~MHz bandwidth divided into four separated sub-bands. Data were converted to filterbank format using a custom-built pipeline \citep{Kirsten2020_nodet} and stored to disk with a 250~kHz frequency and 64~$\mu$s time resolution respectively. Two circular polarisations were averaged together to obtain total intensity data. The telescope has a system equivalent flux density (SEFD) of $458$ Jy at $1.4$ GHz, which leads (using the radiometer equation) to a result of $\sigma_L \simeq 1.2$ Jy ms root mean square (rms) noise for a burst of $1$ ms of duration. To test the data acquisition and conversion we observed PSR B0329+54 and we successfully detected single pulses on this basis.

\begin{figure*}
    \centering
	\includegraphics[width=1.9\columnwidth]{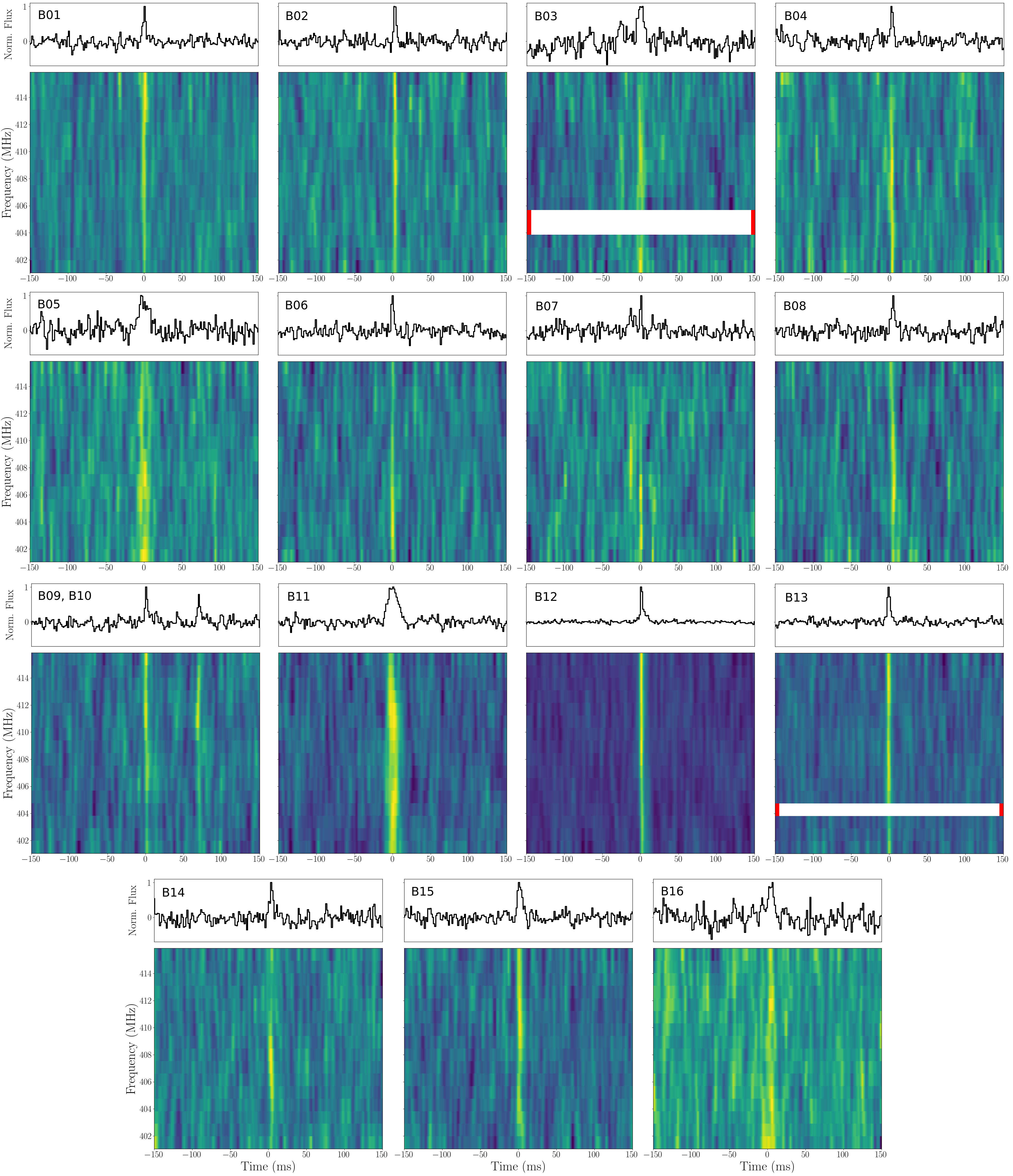}
    \caption{Bursts from FRB 20220912A observed at 408 MHz. Both the de-dispersed, dynamic spectrum (bottom sub-panels) and the frequency-averaged profiles (top sub-panels) are shown. For a better visualisation, the data were down-sampled to have $16$ frequency channels (each $1$ MHz wide) and time bins $1.5$ ms in width. Horizontal white rows (highlighted with red ticks) are flagged channels due to RFI.}
    \label{fig:bursts_12A}
\end{figure*}

\begin{table*}
    \centering
    \caption{Burst properties at 408 MHz from FRB 20220912A.}
    \begin{tabularx}{\textwidth}{l *{7}{c}} 
        \hline
        \hline
        Burst ID  & TOA (B.A.T.) & DM (pc cm$^{-3}$) & Width (ms) & $F_{\rm peak}$ (Jy) & $\mathcal{F}$ (Jy ms) & $E_\nu$ ($10^{30}$ erg Hz$^{-1}$) & $E_{i}$ ($10^{37}$ erg) \\
        \hline
        B01 & 59868.885988115 & $219.6(5)$ & $4.1(5)$ & $18(1)$ & $74(5)$ & $12.3(7)$ & $20(1)$\\
        B02 & 60125.160302085 & $220.2(3)$ & $3.1(4)$ & $7.4(6)$ & $23(2)$ & $3.8(3)$ & $6.1(5)$\\
        B03 & 60133.142754625 & $220.1(2)$ & $9.4(6)$ & $4.3(4)$ & $40(4)$ & $6.7(5)$ & $10.7(8)$\\
        B04 & 60136.132478465 & $219.6(4)$ & $4.1(6)$ & $6.8(6)$ & $28(2)$ & $4.6(4)$ & $7.4(6)$\\
        B05 & 60138.120952875 & $219.7(6)$ & $16(2)$ & $4.8(3)$ & $77(5)$ & $12.8(7)$ & $20(1)$\\
        B06 & 60140.121695755 & $220.4(2)$ & $3.4(4)$ & $7.6(6)$ & $26(2)$ & $4.3(3)$ & $6.9(5)$\\
        B07 & 60160.070367215 & $220.1(2)$ & $1.4(2)$ & $9(1)$ & $13(2)$ & $2.1(3)$ & $3.4(5)$\\
        B08 & 60165.061278865 & $219.4(3)$ & $5.2(8)$ & $7.3(6)$ & $38(3)$ & $6.3(5)$ & $10.1(8)$\\
        B09 & 60166.045907535 & $219.9(2)$ & $3.2(5)$ & $6.2(6)$ & $24(2)$ & $3.3(4)$ & $5.3(6)$\\
        B10 & 60166.045908315 & $219.7(2)$ & $3.1(4)$ & $5.5(7)$ & $20(2)$ & $2.8(3)$ & $4.5(5)$\\
        B11 & 60173.031215725 & $221.3(5)$ & $14.6(8)$ & $10.1(3)$ & $147(5)$ & $24.5(6)$ & $40(1)$\\
        B12 & 60178.024922305 & $220.1(4)$ & $2.9(1)$ & $29.9(7)$ & $86(2)$ & $14.4(3)$ & $23.1(5)$\\
        B13 & 60182.006531445 & $220.2(2)$ & $4.3(3)$ & $13.3(6)$ & $57(3)$ & $9.5(4)$ & $15.2(6)$\\
        B14 & 60211.931989265 & $220(1)$ & $4.3(6)$ & $4.9(5)$ & $21(3)$ & $3.5(4)$ & $5.6(6)$\\
        B15 & 60215.918013385 & $220.8(4)$ & $6.8(7)$ & $6.5(4)$ & $44(3)$ & $7.3(5)$ & $11.8(8)$\\
        B16 & 60218.898463655 & $223(1)$ & $10(2)$ & $3.8(4)$ & $38(5)$ & $6.3(8)$ & $10(1)$\\
        \hline
    \end{tabularx}
    \tablefoot{Columns are, from left to right, the burst ID, the barycentric arrival time (B.A.T) at infinite frequency expressed as the modified Julian day (MJD), the fit-optimized DM, the full width at half maximum (FWHM) duration, the peak flux density, the fluence, the spectral energy and the isotropic burst energy, the latter computed multiplying the spectral energy for $16$ MHz, i.e. the bandwidth used in NC observations.}
    \label{tab:FRB_properties}
\end{table*}

\subsection{uGMRT}

To search for a PRS coincident to the position of FRB 20220912A, continuum radio observations of FRB 20220912A were performed with the uGMRT in the $1050-1450$ MHz (band-$5$) frequency range on UT 2023 November 1$^{\rm st}$. The total bandwidth was splitted into $16384$ channels of $24.414$ kHz each. The field of FRB 20220912A was observed for a total of $\sim 3$ hours. The sources 3C48 and J2322+509, a nearby source to the target, were used as absolute flux scale and phase calibrators, respectively. 

The high spectral resolution of our data enabled us to split the total bandwidth in eight sub-bands of $50$ MHz each for easier data reduction. We processed each sub-band independently by carrying out a standard interferometric data reduction\footnote{See \url{https://science.nrao.edu/facilities/vla/docs/manuals/obsguide/topical-guides/lofreq}} using the Common Astronomy Software Applications \citep[{\sc casa};][]{McMullin07} package. We iteratively performed flagging of RFI, bandpass, amplitude and phase calibrations for each sub-band. Finally, the calibrated visibilities of all the sub-bands were recombined for imaging. We assumed 3C48 to be $17.7$~Jy at 1.2~GHz, with a spectral index $\beta = -0.76$ \citep{PerleyButler13}. These measurements were used for the 3C48 flux and bandpass calibrations, which were then transferred to J2322+509. Finally, we determined gain and phase calibration for J2322+509 and then transferred them to the target field. Owing to severe RFI, two out of eight sub-bands were flagged, thus leaving $300$ MHz of remaining bandwidth. Imaging was carried out with the {\sc tclean} task in {\sc casa}, by weighting the visibilities according to the briggs scheme with a \textsc{robust} parameter $-1$. We achieved a final noise level of $36\ \mu$Jy beam$^{-1}$ at an angular resolution of $1.97'' \times 1.77''$. 

\subsection{Swift and AGILE monitoring}
Since its discovery \citep{McKinven22}, FRB 20220912A has been added to the AGILE list of sources monitored during the Spinning-mode observations. We verified the source exposure with the MiniCalorimeter (MCAL; $0.4\,{\rm MeV}\,\le\,E\le\,100 \,\rm MeV$) detector and the AGILE gamma-ray imaging detector (GRID; 30 MeV\,$\le\, E\, \le\,$ 30 GeV). We took the de-dispersed topocentric arrival times at infinite frequencies as the arrival time for each burst. AGILE acquired MCAL data covering 3 of the 16 bursts presented here, due to the South Atlantic Anomaly (SAA) passages or Earth occultation. We selected good events with standard selection criteria, such as the SAA passages time intervals exclusion, along with the inclusion of events with off-axis angles smaller than $60$ degrees or at angles from Earth direction greater than $80$ degrees.

A monitoring campaign with Swift was also started in autumn 2022, similar to the one dedicated to FRB 20180916B \citep[partially reported in][]{Tavani2020,Trudu23}. Swift observed FRB 20220912A with the X-ray Telescope \citep[XRT, ][]{burrows_2005_swiftxrt}, as one of the three instruments on board. The Swift/XRT X-ray ($0.3$--$10$ keV) data were obtained daily after time of opportunity requests during source activity phases (on November 2022, and July-October, 2023, partially covering the radio monitoring presented in this work). The XRT observations were carried out in windowed timing (WT) readout mode, with $2$--$10$ daily pointings. The time resolution of WT data is $1.8$ ms and each pointing has a typical exposure of $\sim$\,1.8 ks. We considered the combination of all the data and processed them using the XRTDAS software package (v.3.7.0)\footnote{developed by the ASI Space Science Data Center (SSDC)} within the HEASoft package (v.6.32.1). We cleaned and calibrated the data with standard filtering criteria using the xrtpipeline task and the calibration files available from the Swift/XRT CALDB (version 20230705). The imaging analysis was executed selecting events in the energy channels between $0.3$ and $10$ keV and within a $20$ pixel ($\sim\,47''$) radius, including the $90$\% of the point-spread function. The background was estimated from a nearby source-free circular region with the same radius value.
\section{Results and discussion}\label{sec:results}

The search for FRB candidates in NC data follows the strategy employed in \citet{Trudu22} and \cite{Pelliciari23}, using the {\sc spandak} pipeline \citep{Gajjar18}, which flags RFIs through {\sc rfifind} \citep{Ransom2002} and searches for single pulses with {\sc Heimdall} \citep{bbb+12}. We considered a signal-to-noise ratio (S/N) greater than $8$ and a boxcar width shorter than $35$ ms. We carried out our search using DM between $200$ pc cm$^{-3}$ and $240$ pc cm$^{-3}$, given the nominal $220$ pc cm$^{-3}$ DM of the source. In the case of observations at 1.4 GHz, we set the threshold S/N $> 10$, to minimize the RFI contamination. To cross-check the results of the single-pulse search conducted at $1.4$ GHZ with the pipeline described above, we searched for FRBs in a large amount of data also with the processing pipeline described in \cite{Kirsten21, Kirsten23}, which searches FRB with {\sc heimdall} and classifies bursts with the deep learning classifier FETCH \citep{Agarwal20}.We estimated the completeness of our observations at both 408~MHz and 1.4~GHz via injections of simulated burst. We used \textsc{frb-faker}\footnote{\url{https://gitlab.com/houben.ljm/frb-faker}} \citep{Houben19} to inject $100$ bursts of $1$ ms duration with a DM of $220$ pc cm$^{-3}$ at random times. We varied the S/N of the injected burst between $5$ and $20$ and we ran the same pipeline used for the search in order to retrieve the injected bursts. We find a $95\%$ completeness for a fluence of $17$ and $20$~Jy~ms at 408~MHz and 1.4~GHz respectively. 

 \begin{figure*}
 \centering
  \includegraphics[width=1.8\columnwidth]{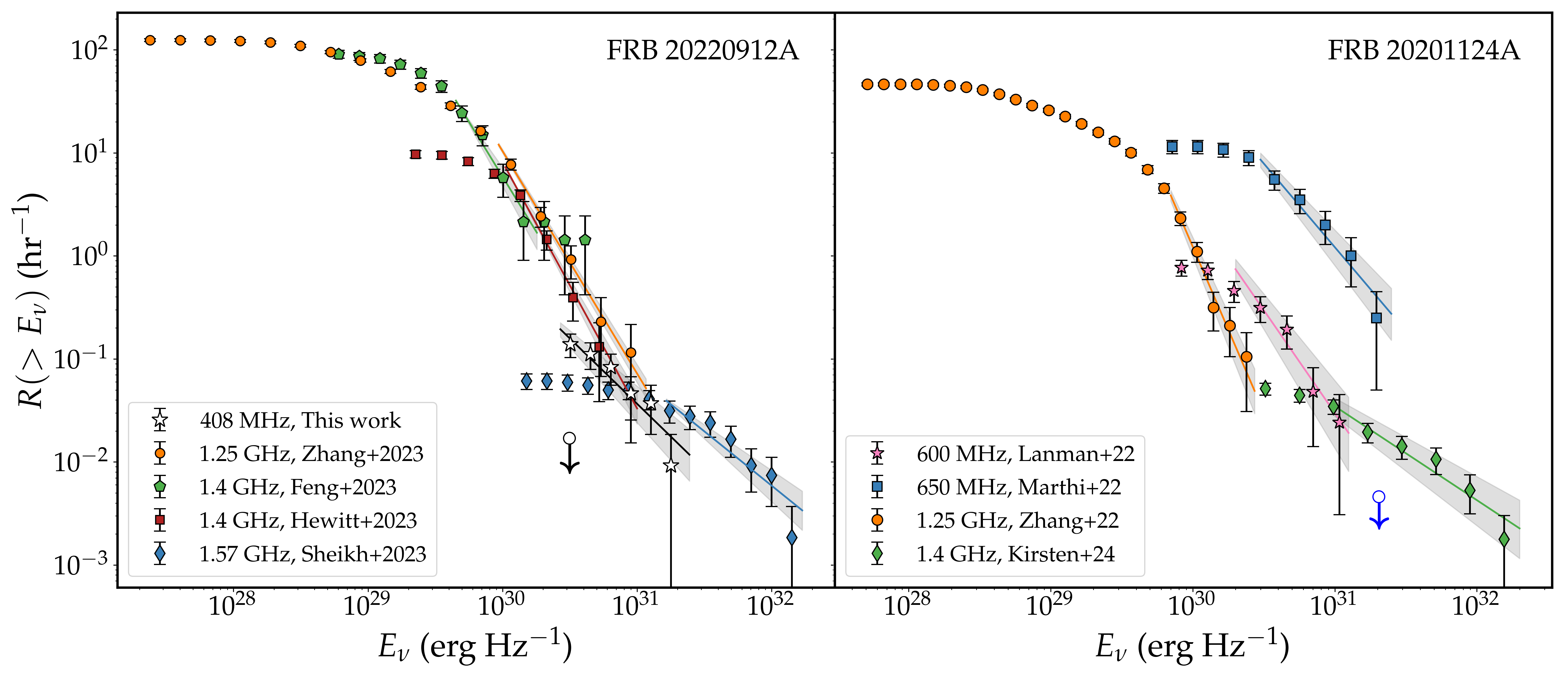}
     \caption{Comparison between the cumulative spectral energy rate distribution for two hyperactive repeaters at different observing frequencies and times. Left: FRB 20220912A at 
    408~MHz (white stars; this work), 1.25~GHz \citep[orange circles;][]{Zhang23}, 1.4~GHz \citep[green pentagons and red squares;][]{Feng23, Hewitt23b} and 1.572~GHz \citep[blue diamonds;][]{Sheikh23}. The $95\%$ C.L. UL for $R (E_\nu > 2.5 \times 10^{30} \, {\rm erg\ Hz}^{-1})$ from our 1.4~GHz observations is plotted as a black downward arrow. Right: FRB 20201124A at $600$ MHz \citep[pink stars;][]{Lanman22}, $650$ MHz \citep[blue squares;][]{Marthi22}, $1.25$ GHz \citep[orange circles;][]{Zhang22c} and $1.4$ GHz \citep[green diamonds;][]{Kirsten23}. The $95\%$ C.L. UL for $R (E_\nu > 2 \times 10^{31} \, {\rm erg\ Hz}^{-1})$ from P ($334$ MHz) resulting from non-detections in \cite{Kirsten23}, is plotted as a blue downward arrows.}
     \label{fig:CED_12A}
\end{figure*}

We detected a total of $16$ bursts at 408 MHz and labeled them as ``Bn'', then ordered them according to their time of arrival (ToA). We show their dynamic, de-dispersed spectra in Fig. \ref{fig:bursts_12A}, while their measured properties are reported in Table \ref{tab:FRB_properties}. All radio bursts, except for the first one, B01, were detected during the second period of the campaign (i.e. later than May $17^{\rm th}$, 2023) when 16 cylinders were used.
We fit a Gaussian profile to the de-dispersed FRB spectrum, integrated over the burst profile, and found that the full width at half maximum (FWHM) of all bursts are compatible with the 16~MHz bandwidth, apart from B14, whose extensions is only $\sim 6$~MHz. Furthermore, we found no evidence of scattering and sub-burst structures on time scales larger than $138.24$~$\mu$s for any of the detected burst. 

Given the $\rm S/N$ of a burst, we computed its fluence as the product of its peak flux density $F_{\rm peak}$ and its FWHM duration $w$, where the former is obtained as \citep{LorimerKramer}:
\begin{equation}\label{eq:radiometer}
    F_{\rm peak} = {\rm S/N} \, \frac{{\rm SEFD}}{A\sqrt{N_p \, N_c \, (1 - \xi) \, \Delta \nu_{\rm ch} \, w}} \, \zeta ({\rm ToA}) \rm .
\end{equation}
Here, $\rm SEFD = 8000$ Jy \citep{Trudu22} holds for each receiver (i.e., each group of sixteen dipoles), $N_p = 1$ is the number of polarisations, $N_c = 1024$ is the number of spectral channels and $\Delta \nu_{\rm ch} = 14.4$ kHz the channel width. Furthermore, $A$ is the number of receivers included in either the eight ($A = 32$) or sixteen ($A = 64$) cylinders, $\xi$ is the fraction of channels excised by RFIs, and, finally, the attenuation of the primary beam at the burst ToA is given by $\zeta(\rm ToA)$ \citep{Trudu22, Pelliciari23}.

Given the burst fluence, we computed the corresponding burst spectral energy $E_{\nu}$:
\begin{equation}\label{eq:energy}
    E_{\rm \nu} = 10^{30}\ \frac{4\pi}{(1+z)^2} \ \Biggl(\frac{D_L}{10^{28}\ {\rm cm}}\Biggr)^2 \ \Biggl(\frac{\mathcal{F}}{{\rm Jy\ ms}}\Biggr)\, \, {\rm erg\ Hz^{-1}}.
\end{equation}
Here the burst isotropic energy is $E_{\rm i}$ \citep[e.g.][]{MacquartEckers18, Chawla22} per unit bandwidth. Also, $D_L = 380.86$~Mpc is the luminosity distance of the source, obtained considering $z = 0.0771$, the redshift of the source, and the Planck 2015 cosmology \citep{Planck15}. 

We computed the cumulative spectral energy rate $R(>E_\nu)$ at 408~MHz. We plot it in Fig.~\ref{fig:CED_12A}, together with spectral energies from literature observations\footnote{Isotropic energies are obtained by multiplying the spectral energies for $16$~MHz at $408$~MHz and $64$~MHz at $1.4$~GHz, respectively.}. It follows a power law in the $3 \times 10^{30} < E_\nu < 3 \times 10^{31}\ {\rm erg\ Hz^{-1}}$ range:
\begin{equation}\label{eq:powerlaw}
    R(> E_\nu) = R_0 (> E_{\nu, 0}) \, \left( \frac{E_\nu}{E_{\nu,0}} \right)^{\alpha_E},
\end{equation}
where $E_{\nu,0} = 3 \times 10^{30}$~erg~Hz$^{-1}$. As a reference, the NC $95\%$ completeness fluence limit corresponds to $3 \times 10^{30}$ erg Hz$^{-1}$. We found the best fit values to be $R_0 = 0.19 \pm 0.03$~hr$^{-1}$ and $\alpha_E = -1.3 \pm 0.2$, respectively. We do not report any detections at $1.4$ GHz, from which we placed a 95$\%$ C.L. upper limit (UL) of $0.017$ hr$^{-1}$ on the burst rate at 1.4~GHz for $\mathcal{F} \geq 20$ Jy ms (corresponding to $E_\nu = 3.2 \times 10^{30}$~erg Hz$^{-1}$).

We repeated the analysis for FRB 20201124A\footnote{We considered a redshift $z = 0.098$ \citep{Kilpatrick21}, corresponding to a luminosity distance $D_L = 453.3$ Mpc \citep{Zhang22c}.}, another very actively repeating FRB source \citep{Xu2022, Zhang22c}. The resulting distributions are shown in Fig. \ref{fig:CED_12A}, while the best fit parameters obtained for each observation are listed in Table \ref{tab:bestfit_energy}.

The cumulative spectral energy rates are fairly similar between FRBs 20220912A and 20201124A, both in the range of the energetic and the repetition rate. This could be an indication that these two sources share the same emission mechanism \citep{James20}, as also highlighted by other similarities such their reported double-peaked waiting time distribution and complex time-frequency structures of their bursts \citep[see e.g.][]{Zhang22c, Zhang23}. Regarding FRB 20220912A, the repetition rate for bursts having  $E_\nu \geq 2 \times 10^{30}$ erg Hz$^{-1}$ decreased from $\sim 10$ hr$^{-1}$ \citep{Zhang23, Feng23} during a storm event, to $\sim 0.1$ hr$^{-1}$, approximately two months later \citep{Sheikh23}, when the storm ended. Lastly, it dropped to less than $0.017$ hr$^{-1}$ nearly a year later, as we could see from our $1.4$ GHz monitoring (see Fig. \ref{fig:CED_12A}). However, the source remained active at $408$ MHz, exhibiting comparable levels of repetition rate to those reported by \cite{Sheikh23}. FRB 20220912A is, as far as we know, the only FRB source that shows a decline of more than four orders of magnitude in its burst rate in the L band.

A similar behaviour, albeit inverted in frequency, can be seen for FRB 20201124A, where the burst rate dropped from $\sim 0.2$ hr$^{-1}$ as reported at $550$ -- $750$ MHz observations \citep{Marthi22} to $<5 \times 10^{-3}$ hr$^{-1}$ at ($2 \times 10^{31}$ erg Hz$^{-1}$, as resulting from non-detections at $334$ MHz in $\sim 650$ hr of observing time \citep{Kirsten23}. To obtain the minimum spectral energy corresponding to the latter observational campaign, we used Eq. \ref{eq:energy}, considering the $91$ Jy ms completeness fluence reported in \cite{Kirsten23}.

Interestingly, we note the same flattening of the cumulative spectral energy rate distribution for high energetic bursts as the one previously reported for R1 \citep{Hewitt22, Jahns23} and, more recently, for FRB 20201124A \citep{Kirsten23}. As can be noted also from the best fit values obtained for $\alpha_E$ in Table \ref{tab:bestfit_energy}, the case for FRB 20220912A is particularly similar to FRB 20201124A. We note that $\alpha_E$ shifts from approximately $-2$ for $E_\nu \geq 10^{30}$ erg Hz$^{-1}$ to roughly $-1$ at $E_\nu \simeq 2 \times 10^{31}$ erg Hz$^{-1}$. Moreover, high energetic bursts present a  slope $\alpha_E = -1.03 \pm 0.3$, obtained by analysing L band data from \cite{Sheikh23}, which is fully consistent with the power law slope, $\alpha_E$, obtained by fitting the cumulative luminosity distribution of apparently non-repeating FRBs \citep{James22b,James22, Shin23}. As a reference, \cite{James22b} obtained $\alpha_E = -0.95^{+0.18}_{-0.15}$, but the other measurements are still consistent with this value. Regarding R1, the cumulative isotropic energy distribution flattens to $\alpha_E = -0.88 \pm 0.01$ for $E_{\rm iso} \geq 1.3 \times 10^{38}$ erg \citep{Jahns23}. This value for $\alpha$ roughly agrees with other reported values as obtained by fitting the R1 cumulative energy distribution at high energies \citep{Law17,Gourdji19,Cruces21,Hewitt22}. Even if the cumulative slopes are similar to the case of FRBs 20220912A and 20201124A, we note that $E_{\rm iso} = 1.3 \times 10^{38}$ erg corresponds to a spectral energy of\footnote{We divided the break istropic energy as reported in \cite{Jahns23} by $450$ MHz, i.e. an average effective bandwidth as reported therein. This value for the break spectral energy agrees well with what reported in \cite{Hewitt22}, when considering a bandwidth of $275$ MHz.} $\sim 4 \times 10^{29}$ erg Hz$^{-1}$, which is approximately one order of magnitude lower than the break spectral energy we obtained for the other two repeaters.

\begin{table*}
	\centering
	\caption{Parameters obtained from the power-law fitting of the cumulative energy distributions for FRBs 20220912A and 20201124A.}	
	\begin{tabular}{lcccc} 
        \hline
        \hline
		 $E_{\nu,0}$ ($10^{30}$ erg Hz$^{-1}$)  & $R_0$ (hr$^{-1}$) & $\alpha_E$ & Ref. \\
		\hline
        \multicolumn{4}{c}{FRB 20220912A}\\
	    1 & $20.5 \pm 3$ & $-2.13 \pm 0.3$ & \citet{Feng23}\\
            1 & $10.5 \pm 1.5$ & $-2.15 \pm 0.01$ & \cite{Zhang23}\\
            2 & $1.51 \pm 0.2$ & $-2.4 \pm 0.2$ & \cite{Hewitt23b}\\
            3 & $0.19 \pm 0.03$ & $-1.3 \pm 0.2$ & This work \\
            25 & $0.036 \pm 0.007$ & $-1.03 \pm 0.3$ & \cite{Sheikh23}\\
            \hline
        \multicolumn{4}{c}{FRB 20201124A}\\
            0.5 & $6.8 \pm 0.5$ & $-2.5 \pm 0.2$ & \cite{Zhang22c} \\
            2 & $0.74 \pm 0.18$ & $-2.0 \pm 0.3$ & \cite{Lanman22} \\
            3 & $9 \pm 2$ & $-1.7 \pm 0.3$ & \cite{Marthi22} \\
            10 & $0.035 \pm 0.01$ & $-0.96 \pm 0.2$ & \cite{Kirsten23}\\
        \hline
	\end{tabular}
 \tablefoot{The first column represents the spectral energy threshold over which data no longer follow a simple power law, the second and third column represent the best-fit values for the parameters of the fitting power-law function (Eq. \ref{eq:powerlaw}). The references for the data that we used to compute the cumulative burst rate distributions are listed in the last column.}
	\label{tab:bestfit_energy}
\end{table*}

\subsection{Constraints on broad-band spectral index}\label{sec:spec_index}

\begin{table}
	\caption{ULs on fluence and broad-band spectral index $\beta$ for Medicina observations for which there is a simultaneous burst detection at $408$ MHz from NC radio telescope.}	
	\begin{tabular}{lccc} 
        \hline
        \hline
		 Burst ID  & $\mathcal{F}_{1.4}$ ($2\sigma$, Jy ms) & $\beta$ \\
		\hline
		B08 & $< 6.3$ & $< -1.6$\\
            B09 & $< 4.9$ & $< -1.4$\\
            B10 & $< 4.8$ & $< -1.1$\\
            B11 & $< 10.1$ & $< -2.3$ \\
            B12 & $< 4.4$ & $< -2.3$ \\
            B13 & $< 5.5$ & $< -1.9$ \\
		\hline
	\end{tabular}
        \tablefoot{The second and third columns represent the limits on fluence and broad-band spectral index, respectively. The former represent the $2\sigma$ detection threshold, where $\sigma$ is the Medicina rms noise computed on a sample of data of duration $\tau$, with $\tau$ the FWHM of the given burst as observed at $408$ MHz.}
	\label{tab:upper_limits}
\end{table}

No multi-band observations of FRB 20220912A have been reported yet. However, from a period of burst storm detected by the Five-hundred-meter Aperture Spherical Telescope (FAST), \cite{Zhang23} obtained a synthetic L band ($1-1.5$ GHz) spectral index of $-2.6 \pm 0.21$ \citep{Zhang23}. This value was obtained by fitting a spectrum obtained by averaging the fluence of all their reported bursts, characterised by having single narrow-band spectra with emission occurring only over $20\%$ of the observing bandwidth \citep{Zhang23}, in different frequency channels. Although this is a valid way to obtain an in-band spectral index, we suggest some caution when making a direct comparison between an UL on the broad-band spectral index and the in-band $\beta$ value obtained in \cite{Zhang23}. Indeed, our observations probe the broad-band spectrum of the source, which can be obtained only when considering simultaneous bursts arriving at separate frequency bands.

During our observational campaign, a total of six bursts (B08-B13) have been detected in the P band during simultaneous L band observations of which we do not report any counterpart. Henceforth, we use these non-detections to provide upper limits on the L band fluence of these bursts, which (in turn) imply ULs on the FRB 20220912A broad-band ($408$ MHz -- $1.4$ GHz) spectral index. For each detected burst with an L band simultaneous observation we computed the fluence UL using the radiometer Eq. \eqref{eq:radiometer}, considering the same width of the corresponding burst in the P band. We considered a $2\sigma$ detection threshold in this case, since our goal has not been to search for bursts blindly with {\sc Heimdall}, which has a minimum $\rm S/N$ search of $\sim 6$. Instead, we manually inspected the Medicina data at burst topocentric arrival times, after correcting them for the DM of the bursts.  The brightest burst we detected during our campaign is B11, with a measured fluence $F = 145 \pm 4.6$ Jy ms at $408$ MHz. We do not report any significant radio emission down to $2\sigma$ at $1.4$ GHz and this translates into a fluence UL of $10.1$ Jy ms in the L band for a burst having a $14.6$ ms duration. This UL translates into $\beta < -2.3$. The same limit on $\beta$ is obtained by the non-detection of an L band counterpart for B12, which has an high S/N as well but with a $\times 10$ shorter duration than B11. We report all the ULs obtained from our observations in Table \ref{tab:upper_limits}.

Our non-detections in the L band undermine the hypothesis of a positive (or flat) broad-band spectral radio emission. We find our UL on $\beta$ in disagreement with $\beta = 2.1$, as measured for a burst from R1 in the multi-frequency, Arecibo (1.4 GHz) -- VLA (3 GHz) campaign; namely, this is the only simultaneous FRB detection present in the literature \citep{Law17}. The other bursts reported in \cite{Law17} have not been detected simultaneously by the two observatories, showing that the broad-band spectral behaviour of the source cannot simply be modelled by a power law function. Moreover, also a spectrum with $\beta \sim 1$ as the one measured from FRB 20200428, the Galactic FRB, simultaneously detected by \cite{CHIME20b} (400 MHz -- 800 MHz) and the Survey for Transient Astronomical Radio Emission 2 \citep[STARE-2, 1.4 GHz][]{Bochenek20a} can be ruled out based on our observations. Therefore, our upper limit could imply either that FRB 20220912A is characterised by a steep radio spectrum or it could be a consequence of its intrinsically narrow-band emission \citep{Zhang23, Feng23, Sheikh23}.  

Our UL is somewhat inconsistent with the very flat spectrum usually observed for radio-loud magnetars \citep{Camilo08, Lazaridis08, Dai19}; however, large fluctuations in the spectral index are observed locally in magnetars, for instance, in the case of XTE J1810--197 \citep{Lazaridis08, Maan22}. An interesting exception is the radio-loud magnetar Swift J1818.0–1607, which has shown emission in a steep spectrum with $\beta \simeq -2.26$ \citep{Lower2022}. A broad-band emission with this spectral index seems to be disfavoured by our observations. Nevertheless, we must be careful in comparing the spectral index for radio-loud magnetars and FRBs, since (up to now) the former showed only pulsed emission \citep[but see also][]{Esposito20}. The only two exceptions to date are SGR J1935+2154 \citep{CHIME20b, Bochenek20a, Zhang20, Kirsten21, CHIME3b} and 1E 1547.0–5408 \citep{Israel21}, which also showed also FRB-like bursts, before entering a pulsar-like phase \citep{Zhu23}. For the former, FRB-like bursts are emitted in random phases (unlike radio pulsations, which instead arrive in a phase windows anti-aligned with X-ray pulsations), hinting at a different emission mechanisms between radio pulses and FRB-like bursts \citep{Zhu23}. For 1E 1547.0–5408, rather, FRB-like bursts are not aligned in phase with radio pulsations, nor with X-ray bursts \citep{Israel21}.

\subsection{Continuum radio emission from FRB 20220912A host galaxy}\label{sec:uGMRT}

\begin{figure}
    \centering
	\includegraphics[width=\columnwidth]{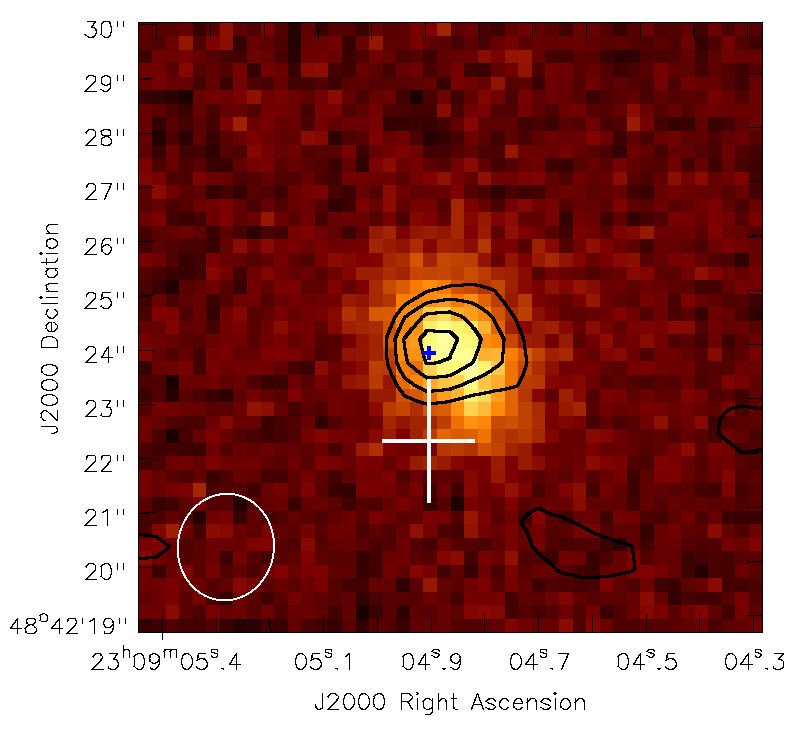}
    \caption{Optical Pan-STARRS (i filter) image of PSO~J347.2702+48.7066, the host galaxy of FRB~20220912A, with contour levels representing the continuum radio source we detected at 1.26~GHz, with a $1.97'' \times 1.77''$ synthesized beam. Contours are
    drawn from three to six times the rms noise level $\sigma \simeq 36$~$\mu$Jy~beam$^{-1}$. The white cross represents the position of APTF J23 radio source, with the cross extensions being the $1\sigma$ uncertainties on its centroid \citep{Hewitt23b}, while the blue cross indicates the position of FRB 20220912A localised at milliarcsecond angular resolution \citep{Hewitt23b}. The synthesized beam of our uGMRT observations is represented in the bottom left corner of the image as a white ellipse.}
    \label{fig:uGMRT_PRS}
\end{figure}

In our uGMRT image at $1.26$ GHz, we detect a continuum source spatially coincident with the coordinates of PSO J347.2702+48.7066, the host galaxy of FRB 20220912A. The source integrated and peak flux densities are consistent at $1\sigma$ level; thus, we considered it as unresolved in our observations. Its centroid have coordinates R.A. (J2000) $=23^{\rm h}09^{\rm m}04.88^{\rm s} \pm 0.017$ s, Dec. (J2000) $=+48^{\circ}42'24.04'' \pm 0.25''$. Such position is well in agreement with R.A. (J2000) $=23^{\rm h}09^{\rm m}04.8988^{\rm s} \pm 0.0003$ s, Dec. (J2000) $=+48^{\circ}42'23.9078'' \pm 0.005''$, namely the localisation of FRB 20220912A obtained from European VLBI Network (EVN) observations \citep{Hewitt23b}. The radio contours of the source we detect are shown in Fig. \ref{fig:uGMRT_PRS}, as well as the VLBI localisation of the FRB. In the same figure, the optical image taken from the Panoramic Survey Telescope \& Rapid Response System (Pan-STARRS) data archive 1 \citep[PS1;][]{Flewelling20} is shown.

We measured a flux density of $240 \pm 36\ \mu$Jy at $1.26$ GHz, which corresponds to a spectral luminosity of $L_\nu \simeq 4 \times 10^{28}$ erg s$^{-1}$ Hz$^{-1}$. We are aware that EVN observations ruled out the presence of a PRS surrounding FRB 20220912A at milliarcsecond scale for an rms of $16$ $\mu$Jy beam$^{-1}$, placing an UL of $1.2 \times 10^{28}$ erg s$^{-1}$ Hz$^{-1}$ on its spectral luminosity at $1.4$ GHz \citep{Hewitt23b}. In the same work, a continuum radio source, APTF J230904+484222 (APTF J23), detected by the  Westerbork Synthesis Radio Telescope Aperture Tile In Focus (WSRT-APERTIF) has been reported, with a position consistent with the coordinates of FRB 20220912A host galaxy. This source has a peak flux density of $270 \pm 40\ \mu$Jy beam$^{-1}$, which falls within the measured flux range of our source, considering the associated uncertainties. Although APTF J23 is offset by $\sim 1.6''$ with respect to the FRB VLBI position (see Fig. \ref{fig:uGMRT_PRS}), the $\sim 2''$ uncertainties associated to its centroid makes APTF J23 consistent at $2\sigma$ level with the position of our detected source. We thus conclude that the source we detect and APTF J23 are the same radio source, albeit observed in this work with an improved spatial resolution by a factor of four.

Finally, we note that the contour levels of the persistent radio source we are aiming to detect are offset by approximately $0.6''$ (about $0.9$ kpc in physical size) from the geometric centre of the host galaxy. This offset lowers the plausibility that an active galactic nucleus (AGN) is the origin of this source and, instead, this suggests that the radio emission is originating from a star formation region in the vicinity of FRB 20220912A.

\begin{table*}
\caption[AGILE burst coverage and fluence ULs]{\emph{AGILE} FRB 20220912A bursts coverage and MCAL ULs.}
\label{tab:AGILE_FRB}
\centering                          
\begin{tabular}{l c c c c c c}        
\hline\hline                 
Burst & MCAL & MCAL & GRID  & UL & MCAL & UL \\
   ID & FoV & D.A. & FoV &  ($3\sigma$) & trigger FAR & 1 ms \\
      & & & & $\mathrm{[erg \ cm^{-2} ] }$ & [evt/hour] & $\mathrm{[erg \ cm^{-2} ] }$ \\
\hline
B01 &  \multicolumn{6}{c}{idle mode} \\
B02 & YES & NO & NO   & -- & -- & $3.22 \times 10^{-8}$\\
B03 &  \multicolumn{6}{c}{idle mode} \\
B04 & YES & NO & NO  & -- & -- & $2.59 \times 10^{-8}$\\
B05 & YES & NO & YES   & -- & -- & $1.84 \times 10^{-8}$\\
B06 & YES & NO & NO   & -- & -- & $5.15 \times 10^{-8}$\\
B07 &  \multicolumn{6}{c}{no data} \\
B08 & YES & NO & NO   & -- & -- & $4.22\times 10^{-8}$\\
B09 & YES & YES & NO   & $2.06 \times 10^{-7}$ & $\sim 4.0$ & --\\
B10 & YES & YES & NO   & $2.04 \times 10^{-7}$ & $\sim 4.0$ & --\\
B11 & YES & YES & NO   & $2.00 \times 10^{-7}$ & $\sim 4.0$ & --\\
B12 & YES & NO & NO   & -- & -- & $2.17 \times 10^{-8}$\\
B13 &  \multicolumn{6}{c}{idle mode} \\
B14 &  \multicolumn{6}{c}{idle mode} \\
B15 & YES & NO & YES   & -- & -- & $1.83 \times 10^{-8}$\\
B16 &  \multicolumn{6}{c}{no data} \\
\hline 
\end{tabular}
\tablefoot{The second and fourth columns report the presence or absence of the source in the field of view (FoV) of the two onboard detectors here considered, respectively. 
In the third column we report the existence of an MCAL data acquisition at trigger time.
MCAL fluence $3\sigma$ ULs in 0.4\,--\,30 MeV band are evaluated only when no coincident data acquisitions at the burst times is present; MCAL trigger "False Alarm Rate" (FAR) are evaluated when coincident data acquisition is present. ULs (1 ms) refer to the UL fluences that would be required to issue a trigger with the onboard 1 ms MCAL trigger logic timescale \citep[see][]{2022ApJ...924...80U}. Since there is no evidence of a detection in the acquired triggers, we  checked in the 100 days preceding each burst time to estimate the FAR corresponding to each trigger.}
\end{table*}

\begin{table*}
\centering
\caption{Swift exposures and flux ULs (0.3\,--\,10 keV).}
\label{tab:swiftcov}
\begin{tabular}{c c c }       
\hline\hline                 
Start time  & Stop time  &  UL  \\
        \multicolumn{2}{c}{(UTC)}
&  $\mathrm{[erg\; cm^{-2}\, s^{-1}]}$ \\
\hline
2022-11-11 19:26:15   & 2022-11-20 18:29:56 & 2.3 $\times 10^{-13}$ \\
2023-07-25 00:09:32 & 2023-07-30 02:47:55 & 2.5 $\times 10^{-13}$ \\
2023-08-29 21:26:07 & 2023-09-09 21:01:56 & 2.1 $\times 10^{-13}$ \\  
2023-09-29 22:27:14 & 2023-10-05 21:30:56 & 4.9 $\times 10^{-13}$ \\   
\hline 
\end{tabular}
\end{table*}

\subsection{Results from the high energy monitoring of FRB 20220912A}
\label{sec:obs_xgamma}
As a first step, we checked the burst exposures in the AGILE source monitoring, as well as the position of the source within the AGILE FoV at each burst time. The relative AGILE exposure to the source is reported in Table \ref{tab:AGILE_FRB}. We obtained a coverage of three of the observed bursts with MCAL data but no detection was found analyzing the light curves in five binnings (16, 32, 64, 256 ms, and 1 s) and considering shifts of 1/4 of bin (four shifts for the first two time scales, two for the second two).
We extracted a $3\sigma$ C.L. fluence ULs in $0.4$ -- $30$ MeV energy band considering a cut-off power-law model, with a photon index of $-0.70$ and cut-off energy of 65 keV \citep[as reported for the FRB 200428 burst by] [] {Mereghetti20}. We also estimated UL fluences that would be required to issue a trigger with the onboard $1$ ms MCAL trigger logic timescale \citep[see][]{Ursi22}. We report the corresponding ULs in Table \ref{tab:AGILE_FRB}. These MCAL UL values are somewhat lower than the similar values previously published \citep[see for instance][]{Trudu23}, thanks to the non-standard spectral model applied in this work.
The most stringent UL that we can place on the radio efficiency $\eta = E_{\gamma}/E_{\rm Radio}$ is from burst B05, for which the non-activation of the MCAL trigger system permits us to obtain $\eta < 1.5 \times 10^9$ at $3\sigma$ C.L. in the $0.4$ -- $30$ MeV energy range. This UL is more stringent than what found from the radio non-detection of a giant flare from magnetar SGR 1806-20, as already reported also for FRB20180916B in \cite{Tavani2020,Tavani21}, for which $\eta \sim 10^{11}$ \citep{tendulkar16}. 
We conclude that observations in this work confirm the exclusion of giant X-ray flares as possible X-ray countepart of B05. Moreover, we set an UL value on $\eta$ consistent with those previously reported for one-off and repeater sources \citep[see, e.g., Figure 3 from][and references therein]{Pearlman23}.
We note that our UL for $\eta$ is conservative. Indeed, $\eta$ depends on the radio isotropic energy of the burst, which in turn relies on the spectral occupancy. In our case, we considered $\Delta \nu = 16$ MHz, which is the observed bandwidth of the NC radio telescope, but the unknown intrinsic spectral width of the burst likely exceeds this value, and it could allow tighter constraints to be placed on $\eta$. 

AGILE/GRID covered 2 of the 16 bursts (B05 and B15) at their ToA. We analyzed GRID data near burst arrival times on short ($± 100$ s around the bursts), and longer timescales ($\pm 10$ days and $100$ days starting about B05 trigger time). The long-timescale data analysis was performed applying the standard AGILE multi-source maximum likelihood \citep[AML;][]{Bulgarelli2012}, which is mainly applied to exposures longer than a few hours. We report no detection at short timescales for AGILE/GRID. Finally, we extracted 3\,$\sigma$ ULs in the E\,$\geq\,$100\,MeV band for two long time integrations, $10$ days after each burst and $100$ days after burst B05 (the latter period of time includes also B15). We obtained UL$_{10d}$\,=\,2.0\,$\times\,10^{-11}$ \,erg\,cm$^{-2}$\,s$^{-1}$ and UL$_{100d}$\,=\,4.4\,$\times\,10^{-12}$\,erg\,cm$^{-2}$\,s$^{-1}$. From the latter, we obtained $L_\gamma < 7.1 \times 10^{43}$ erg s$^{-1}$ for the persistent $\gamma$-ray luminosity of the source.

No X-ray source was detected at $>3\sigma$ C.L. in the whole Swift/XRT WT mode dataset. We note however, with a detailed single observation check, that no radio burst was exposed even including three more proposals acquired in photon counting (PC) mode: burst B01 occurred 16 hrs after the observation on October 16th, 2022, while B16 occurred within our third observations but did not fall within the WT mode sky window. We then extracted a 3\,$\sigma$ countrate ULs for our observations using the XIMAGE package (sosta command) and converted to fluxes using a standard single power-law spectral model with a photon index of $2.0$, and correcting for absorption for a column density of N$_H$ fixed to the Galactic value of 1.43\,$\times\,10^{21} \rm cm^{-2}$ \citep{HI4PI_2016} redshifted for the redshift of the source \citep{ravi23}. The X-ray observations exposure and the corresponding ULs for the persistent X-ray fluence are reported in Table~\ref{tab:swiftcov}. From the latter, we obtained an UL for the persistent X-ray luminosity $L_X = 4\pi D_L^2 F_X / (1+z) < 3.4 \times 10^{42}$ erg s$^{-1}$, where $D_L$ is the luminosity distance of FRB 20220912A. 
We note that this UL excludes the majority of mid- and high-luminosity AGNs, which typically have X-ray luminosity $L_X \geq 10^{43}$ erg s$^{-1}$ \citep[e.g.][]{Padovani17}. This offers further evidence that the radio source detected with uGMRT (see Section \ref{sec:uGMRT}) originates from a region of star formation, rather than from an AGN.

\section{Summary and conclusions}\label{sec:conclusions}

In this work, we present a campaign of simultaneous observations at 408 MHz and 1.4 GHz, taken with the NC and Mc radiotelescopes, respectively, of one of the most active repeaters known to date: FRB 20220912A. During the campaign, we detected 16 bursts from FRB 20220912A at $408$ MHz. We found that the cumulative burst rate as a function of the spectral energy at $408$ MHz can be aptly fitted with a single power-law function $R(>E_\nu) \propto E_\nu^{\alpha_E}$, with $\alpha_E = -1.3 \pm 0.2$. We do not report any burst detection at $1.4$ GHz in a total of $177$ hr above a fluence and a spectral energy threshold of $\mathcal{F} \geq 20$ Jy ms and $3.2 \times 10^{30}$ erg Hz$^{-1}$, respectively. These non-detections place an UL of $0.017$ hr$^{-1}$ at $95\%$ C.L for the burst rate at $1.4$ GHz, which is about four orders of magnitude lower than the level of activity reported at the same frequency during a burst storm of the source \citep{Zhang23, Feng23} at the same spectral energy. On the other hand, the source remained active at $408$ MHz with comparable repetition rate as observed from a long monitoring after the end of the burst storm \citep{Sheikh23}. Interestingly, we note that the cumulative spectral energy rate distribution of FRB 20220912A flattens for bursts having a spectral energy of $E_\nu \geq 10^{31}$ erg Hz$^{-1}$, changing slope from approximately $\alpha_E \approx -2$ to $\alpha_E \approx -1$. This flattening feature has been reported so far for two other well-studied hyperactive repeaters, R1 \citep{Hewitt22, Jahns23} and FRB 20201124A \citep{Kirsten23}. As discussed in \cite{Kirsten23}, this could be linked to a different type of emission mechanism, emission site or beaming angle between low and high energy bursts. This could potentially represent a link between one-off bursts and repeating sources \citep{James22,Kirsten23}. The fact that also FRB 20220912A shows this kind of behaviour, combined with the fact that high energy bursts present a slope that is amply consistent with that of the non-repeaters population \citep{James22b, James22, Shin23}, provides further support to the idea that a fraction of apparently non-repeating FRBs could (instead) be repeating sources with very low repetition rates \citep{Ravi19, James23}. We compared the cumulative spectral energy rate for FRB 20220912A and FRB 20201124A, highlighting a strong similarity between the two distributions, both in terms of the spectral energy and repetition rate ranges.

In total, $6$ of the $16$ bursts detected at $408$ MHz arrived during simultaneous observations at $1.4$ GHz, allowing us to place the first ULs to the broad-band ($408$ MHz -- $1.4$ GHz) spectral index $\beta$ of FRB 20220912A. Analysing the Mc data at the ToA of burst B11, which has the highest fluence, we obtained an UL of $\beta < -2.3$, indicating that the source (under the assumption of an intrinsically broad-band emission) exhibits a very steep spectral index. Our observations then strongly disfavor a flat or even inverted spectrum for FRB 20220912A. We note that this is different than what has been reported for the only two simultaneous detections of FRBs in separate frequency bands, the latter indicating a positive spectral index \citep{Law17, Bochenek20a, CHIME20b}. Our findings support the idea that the intrinsic spectrum of FRB 20220912A's bursts is narrow-band, as reported in recent observations \citep{Zhang23, Feng23, Sheikh23, Hewitt23b}.

Additionally, we reported three-hour-long continuum radio observations of FRB 20220912A field using the band 5 ($1050-1450$ MHz) of the uGMRT. We detected a continuum radio source of $240 \pm 40\ \mu$Jy flux density that is spatially coincident with the FRB 20220912A VLBI localisation. Given it has not been detected in recent, deep, EVN observations \citep{Hewitt23b}, we suggest that this continuum radio source may possibly originate from a region of star formation, potentially located in the vicinity of the FRB source. This is corroborated by the $0.6''$ offset (0.9 kpc in physical size) between the source centroid and the geometric centre of the optical host galaxy, which excludes the hypothesis that the radio source is being powered by an AGN.

Finally, we report the results of an X- and $\gamma$-ray monitoring of FRB 20220912A with the Swift and AGILE space missions in X- and $\gamma$ rays, respectively. We reported no detection for either of these high-energy campaigns. Regarding AGILE, we placed an UL on the radio efficiency of $\eta = E_{\gamma} / E_{\rm radio} < 1.5 \times 10^9$ for the B05 burst, the latter being more stringent than the UL on $\eta$ obtained from the radio non-detection of a giant flare from magnetar SGR 1806-20 \citep{tendulkar16} and consistent with literature ULs from one-off sources and repeaters \citep{Pearlman23}, along with a persistent $\gamma$-ray luminosity UL of $L_\gamma < 7.1 \times 10^{43}$ erg s$^{-1}$. From Swift observations we obtained instead a UL for the persistent X-ray luminosity (0.3 -- 10 keV) of $L_X < 3.4 \times 10^{42}$ erg s$^{-1}$.

\begin{acknowledgements}
We thank the anonymous referee for the useful comments, helping us in improving the quality of the paper. The reported data were collected during the phase of the INAF scientific exploitation with the NC radio telescope. This article was produced with the support of the PhD program in Space Science and Technology at the University of Trento, Cycle XXXIX, with the support of a scholarship financed by the Ministerial Decree no. 118 of 2nd March 2023, based on the NRRP - funded by the European Union - NextGenerationEU - Mission 4 "Education and Research", Component 1 "Enhancement of the offer of educational services: from nurseries to universities” - Investment 4.1 “Extension of the number of research doctorates and innovative doctorates for public administration and cultural heritage” - CUP E66E23000110001 and of the Scuola Universitaria Superiore IUSS Pavia.
\end{acknowledgements}

\bibliography{biblio}{}

\bibliographystyle{aasjournal}

\end{document}